\documentclass[11pt]{amsart}
\usepackage{geometry}                
\usepackage{graphicx}
\usepackage{amssymb}
\usepackage{epstopdf}
\usepackage{amsmath}
\def\pre#1#2#3{{\it Phys. Rev. E} {\bf #1}, #2 (#3)}
\def\CHAOS{{\it CHAOS }}
\def\X{{\bf{X}}}

\def\eg{e.g.~}
\def\etl{$et ~al.$}

\def\iota{i}

\begin{document}

\title{Amplitude Death: The cessation of oscillations in coupled nonlinear dynamical systems}
\author{Garima Saxena, Nirmal Punetha, Awadhesh Prasad, Ram Ramaswamy}

\begin{abstract}
Here we extend a recent review (Physics Reports {\bf 521}, 205 (2012)) of amplitude death, namely the suppression of oscillations due to the coupling interactions between nonlinear dynamical systems. This is an important emergent phenomenon that is operative under a variety of scenarios.  We summarize results of recent studies that have significantly added to our understanding of the mechanisms that underlie the process, and also discuss the phase--flip transition, a characteristic and unusual effect that occurs in the transient dynamics as the oscillations die out. 
\end{abstract}

\maketitle

Nonlinear systems can show a range of complex dynamics  depending on the nature of the 
equations of motion. When two or more systems are coupled, then there is frequently newer, emergent behavior that depends on the manner in which the systems interact.  Synchronization is one such phenomenon, but  depending upon the manner in which the coupling is organized, the collective dynamics can be more complex \cite{hys,rid}. An unusual and unexpected consequence of the coupling is to induce simplicity: in addition to synchrony \cite{kurth}, there can be the suppression of chaos and appearance of periodicity  \cite{rajatphyscon}, and amplitude death (AD), namely the loss of any oscillatory dynamics when the dynamics is driven to a fixed point \cite{review}.  

In the past few decades, AD has been the subject of extensive study due to 
potential applications in stabilizing systems to the steady state.
Oscillation quenching is often desired as a control mechanism in technology: 
for suppressing fluctuations in the power output of lasers \cite{laser}, 
in thermo--optical oscillators for implementing safety measures \cite{thermo}, 
in coupled self excited elastic beans \cite{mech}, or 
in electrical engineering to stabilize DC grids with constant power loads \cite{power} for instance, 
and also for medical purposes like treating neuronal disorders \cite{nd1,nd2,nd3}. In other applications,
AD has also been proposed as an underlying mechanism for auditory transduction \cite{hear} and
is also presumed to play an important role in climatology, where the large scale oceanic and atmospheric anomalies are found to be  correlated with the zonal coupling of atmospheres of the respective ocean basins \cite{ocean}. 

A recent review  \cite{review} has focussed on AD in different fields. The characteristics of
different coupling  strategies and scenarios that lead to AD and its
occurrence in networks of coupled oscillators and in various experimental situations has been
discussed in detail.  We summarize these briefly here. Starting with the work by Aronson \etl \cite{aronson} who showed that mismatched units, when coupled lead to AD, several other scenarios have been proposed. Time--delay coupling \cite{delay1,delay2} and conjugate coupling \cite{cc} 
both cause identical coupled systems to show AD. 

In configurations where the parameters of individual systems are not accessible, a strategy that has been proposed is the so--called dynamic coupling \cite{dc}, where  the coupling variable itself has nontrivial dynamics in the absence of interaction. This coupling scheme has been explored in a variety of different system configurations \cite{dcall} and been shown to be a robust mechanism to cause AD.   All these configurations work with linear coupling, and by taking the coupling itself to be nonlinear, one can, in addition, stabilize a targeted or ``designed'' steady state 
\cite{nlc}. AD can also be achieved by coupling a linear and nonlinear system \cite{la}, in a configuration termed  {\it linear augmentation}.  
When the coupling involves transmission delay, AD follows as a natural consequence \cite{delay1,delay2}. More realistic situations are modeled by including delays that are not fixed but are distributed in some manner and this is shown to enhance the region of stability in the parameter space  \cite{atay,garima}. 

Recently, Konishi and coworkers \cite{konishi} have shown that AD can be realized more efficiently when the delay is itself time--varying. Other schemes such as environmental coupling \cite{amritkar}, mean--field diffusion \cite{meanf},  gradient coupling \cite{grad}, partial time varying delay \cite{partial} 
and indirect coupling \cite{indirect} for stabilizing AD have also been proposed. 
In this review, in the next section we discuss these recently proposed scenarios for AD, namely time--varying delays, environmental coupling and other types of interactions. 
The {\it phase-flip} transition is discussed in Section 2, and AD in coupled Hamiltonian systems in Section 3.  We conclude with a brief summary in Section 4.

\section{Scenarios }

For consistency we discuss the various scenarios for AD in the context of coupled Landau--Stuart limit cycle oscillators. The equation of motion for a single Landau--Stuart oscillator is  
\begin{equation} 
\dot{Z}(t)=(A+i\omega+|Z(t)|^{2})Z(t) 
\label{eq:ls}
\end{equation}
where, $Z \equiv x+iy$ is a complex variable, $A$ determines the degree of instability of the fixed point $Z^{*}=0$ and $\omega$ is 
the natural frequency of oscillations. 

\subsection{Variable time--delay coupling}
It is well known that AD can occur when identical systems are coupled with time--delay. In such cases there are two additional parameters apart from those of the 
individual system, namely the coupling strength $\epsilon$ and the time delay $\tau$. Depending upon the system properties, AD occurs in regions  
in parameter space $(\epsilon,\tau)$; these appear as distinct `death islands'  \cite{delay1}, so it would
seem that AD can be realized practically for only a limited range of $\tau$. 
In practical situations, short delay times are not easily implemented, and this can pose a  problem.
A recent approach to addressing this limitation has been to make the time--delay  itself time dependent, and this is easily implemented in experiments \cite{konishi2}. 

Consider a pair of Landau--Stuart oscillators,
\begin{eqnarray}
\dot{Z}_{1}(t)=(A+i\omega+|Z(t)|^{2})Z_{1}(t)+u_{1}(t) \nonumber \\
\dot{Z}_{2}(t)=(A+i\omega+|Z(t)|^{2})Z_{2}(t)+u_{2}(t) 
\label{eq:tv}
\end{eqnarray}
where the coupling $u_i(t)$ is 
\begin{eqnarray}
u_{1,2}(t)=\epsilon [Z_{2,1}(t-\tau (t))-Z_{1,2}(t)]. \nonumber
\end{eqnarray}
The time dependent time--delay $\tau (t) \geq 0$ in the coupling signal can be chosen such that it varies periodically around an average,  $\tau_{0}$, 
\begin{eqnarray}
\tau(t):=\tau_{0}+\delta f(\Omega t),  \nonumber
\label{eq:tvd}
\end{eqnarray}
 and Konishi {\it et al.}  \cite{konishi} took $f(\Omega t)$ to be a sawtooth function, 
\begin{eqnarray} 
f(x):= +\frac{2x}{\pi}-1-4m & \mbox{if} & x\in[2m\pi, (2m+1)\pi] \nonumber \\
       -\frac{2x}{\pi}+3+4m & \mbox{if} & x\in[(2m+1)\pi,2(m+1)\pi]  \nonumber
\label{eq:tvd-shaw}
\end{eqnarray} 
for $m$ = 0, 1, 2.... Note that $\delta \in [0,\tau_{0}]$ and $\Omega >0$ are the amplitude and frequency of variation.  

\begin{figure}
\centering
\includegraphics[width=8cm]{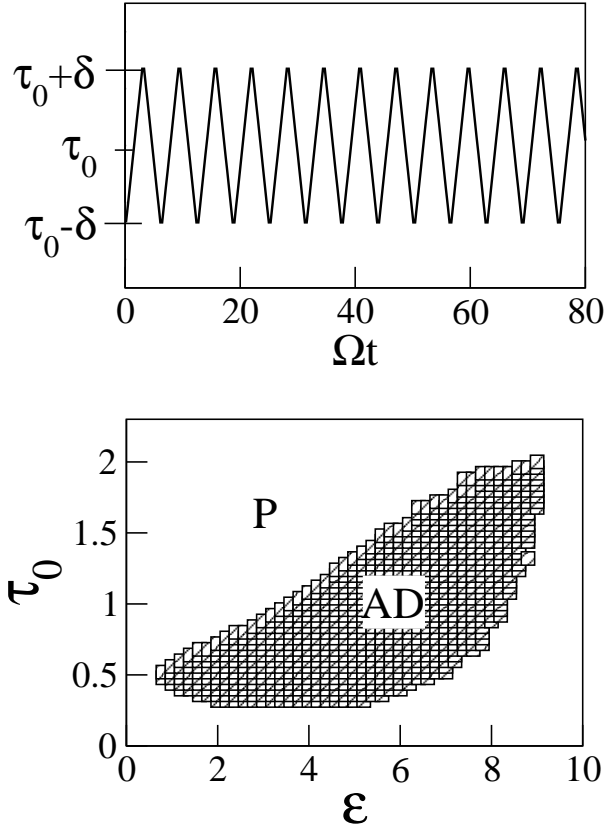}
\vskip1cm
\caption{\label{fig:tvdelay} (a) The variation of time-delay $\tau(t)$ as a function of time, for $\tau_{0}=1$,   Eq.(\ref{eq:tvd-shaw}), 
(b) The schematic phase diagram in parameter space $(\epsilon-\tau_0)$ for $\delta=0.3$. AD occurs in the shaded region.}
\label{fig:pspace}
\end{figure}

In numerical simulations the parameters are taken as $A$=0.5, $\omega=\pi$ and $\delta=0.3$ in \cite{konishi}. The variation of $\tau(t)$ is shown in Fig.~\ref{fig:tvdelay}(a) and the dynamics of the coupled system as a function of the parameters $(\epsilon, \tau)$ in Fig.~\ref{fig:pspace}(b). The dotted region corresponds to AD where the fixed point $Z_i^*=0$ is stabilized and elsewhere the dynamics is periodic. When the amplitude $\delta$ is increased, the AD region in parameter space increases correspondingly. Konishi \etl \cite{konishi} derive the analytic conditions for AD for arbitrary time--delay values in the above system to obtain the result that when $\delta=\pi/\omega$ where $\omega<<\Omega$ and $\mu<\omega(2+\pi)/4\pi$, the AD region becomes unbounded. Thus the use of time--dependent delay provides a systematic procedure for designing AD by tuning appropriate system parameters. 

\subsection{Environmental Coupling}

When the interaction between two systems is mediated through an external agency, namely an   
{\it environmental} effect  \cite{amritkar}, a number of interesting phenomena arise. Such indirect or ``relay'' coupling is quite common: genetic oscillators, for instance, are typically coupled via an external agency  such as the cell membrane through which chemical diffusion occurs.  Examples can be  drawn from a number of fields  that ranging from chemical oscillators to atomic ensembles where coupling is  implemented by the surrounding media \cite{amritkar}.

One model for the dynamics of a system that is coupled through the environment is given by the equations of motion
\begin{eqnarray}
\dot{\X}&=&f(\X)+\beta s \nonumber \\
\dot{s}&=&g(s)+h(\beta, \X) \nonumber
\end{eqnarray}\noindent
where $\X \in \mathbb{R}^{m}$ are the variables of the dynamical system and $s \in \mathbb{R}^{1}$ represents the action of the environment. The vector $\beta$ with elements 0 or 1 selects components of $\X$ that are affected by the environment. 
It has been shown by Amritkar and coworkers \cite{amritkar} that this form of environmental coupling can induce AD in the coupled units that also interact with one another directly so as be synchronized in--phase. Environmental coupling in effect frustrates the system by giving each unit an anti--phase synchronizing tendency. This forces the synchronized system to cease oscillating when both couplings (one causing in--phase synchronization and second inducing anti--phase synchronization) are operational and there is a competition between the two tendencies, leading to AD. 
This mechanism does not stabilize the unstable fixed point of the uncoupled systems; the new interactions generate novel steady states so that this form of behavior is more properly termed oscillation death.   We illustrate this in the coupled Landau--Stuart system
\begin{eqnarray}
\dot{Z}_{1}(t)&=&(1+i\omega+|Z_{1}(t)|^{2})Z_{1}(t)+\epsilon_{1}[Z_{2}(t)-Z_{1}(t)]+\epsilon_{2}s  \nonumber \\
\dot{Z}_{2}(t)&=&(1+i\omega+|Z_{2}(t)|^{2})Z_{2}(t)+\epsilon_{1}[Z_{1}(t)-Z_{2}(t)]+\epsilon_{2}s  \\
\dot{s}(t)&=&-s-\frac{\epsilon_{2}}{4} \sum (x_{i}+y_{i}); \nonumber
\label{eq:envc}
\end{eqnarray}
where $\beta$ is a vector of length 4 with all elements taken to be unity.  $\epsilon_2$ is the environmental coupling strength, $\epsilon_1$ that of the direct coupling, and the frequency $\omega$ is taken here to be 10.

\begin{figure}
\centering
\includegraphics[width=10cm]{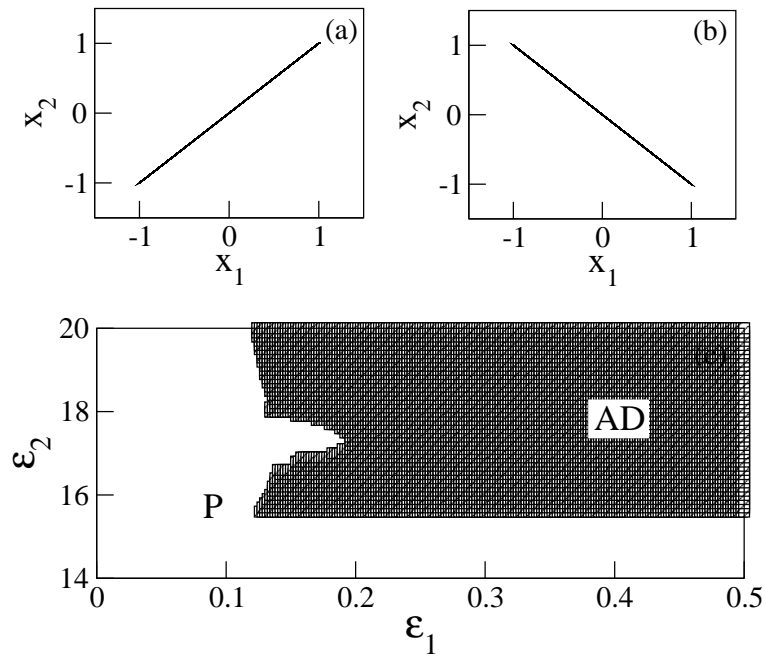}
\vskip.5cm
\caption{(a) In--phase synchronization for $(\epsilon_{1},\epsilon_{2})=(0.2,0)$ and 
(b) anti--phase synchronization when $(\epsilon_{1},\epsilon_{2})=(0,0.2)$. (c) A schematic phase diagram in  parameter space $(\epsilon_{1},\epsilon_{2})$. AD occurs in the shaded region, outside which the dynamics is oscillatory.}
\label{fig:env1}
\end{figure}

Fig.~\ref{fig:env1}a corresponds to the case when the environment is switched off, namely $\epsilon_{2}=0$ and the oscillations synchronize. Similarly, when only environment coupling is present, namely $\epsilon_{1}$ = 0, the oscillators are in anti--phase synchronization, Fig. \ref{fig:env1}b. When both diffusive and environmental coupling are switched on the oscillators are driven to AD, and the region where such behavior is shown as a function of the two coupling parameters $\epsilon_1,\epsilon_{2}$ in  Fig.~\ref{fig:env1}(c). 

While here the environment has been  modeled as an over-damped oscillator that is kept active through feedback from the dynamical system(s) \cite{amritkar2} AD can occur even when the environment has other intrinsic dynamics. Further, such stabilization also works in networks of nonlinear oscillators \cite{amritkar3}.

\subsection{Other Strategies}
We summarize some recent studies that have proposed methods causing AD that are variants or extensions  of the scenarios discussed in Ref. \cite{review}.
 
{\it Mean Field Diffusion} \cite{meanf} which induces AD in identical systems coupled 
by similar variables. The studies so far showed that AD in such a configuration can occur only under 
parameter mismatch \cite{aronson}. Mean field interaction releases this constraint by introducing 
a control parameter in the coupling and tuning it within the optimal range. 
It is a modification to the widely studied diffusive coupling \cite{diffc}.
Though the name of the coupling suggests its existence in networks, it is also true for the limiting case 
of two coupled systems.

\begin{figure}
\centering
\includegraphics[width=8cm]{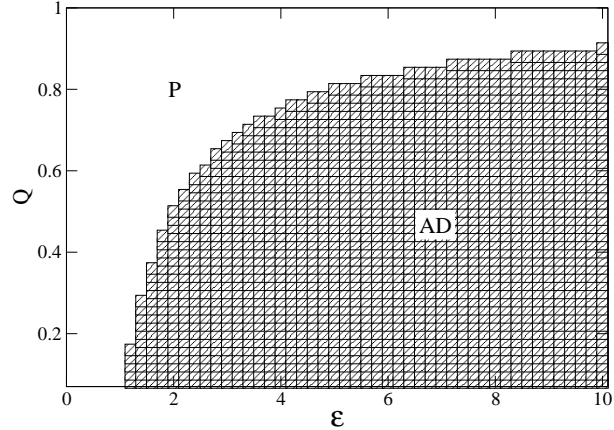}
\caption{The schematic phase diagram in parameter space $(Q,\epsilon)$. AD occurs in the shaded region.}
\label{fig:meanfield}
\end{figure}

Consider $N$ coupled Landau--Stuart oscillators, 
\begin{eqnarray}
\dot{Z_{i}(t)} &=& (\mu+i\omega_{i}+|Z_{i}|^{2})Z_{i}(t))+\epsilon(Q\overline{Z}-Z_{i}) \\
\overline{Z}&=&\frac{1}{N}\sum_{i=1}^N Z_{i} \nonumber
\label{eqn:mf}
\end{eqnarray}
with $Q$ the mean--field control parameter that determines the extent of feedback. 
In the limit $Q \to 0$ when the oscillators decouple,  and $Q \to $ 1, 
that maximizes the interaction, there is no AD but for some intermediate $Q$ there can be oscillation death.  For $ Q < 1$ the effect of mean field is reduced, causing the limit cycles to pull each other towards the steady state; this can be seen in the phase diagram plotted in 
Fig. ~\ref{fig:meanfield} for $N$=2.  AD also occurs when $Q$ is made time dependent \cite{ap}.

A scenario that has been studied recently highlights the significance of asymmetrical coupling.  
Asymmetry can arise due to different coupling strengths by which systems interact with each other, reflected in the different 
$\epsilon$ values for the subsystems. It induces AD in identical coupled systems over a substantially large regime in coupling parameter space 
\cite{asym1}. In the presence of delay it further enhances the region in which AD occurs \cite{asymeps}.
Asymmetry also arises when attractive $(\epsilon > 0)$ and repulsive $(\epsilon < 0)$ coupling are together present \cite{rep}. This scenario shows 
rich dynamical behavior along with the occurrence of AD.

Other studies of AD have examined the combined effect of parameter mismatch and delay \cite{orderpara}, 
designing AD by single \cite{sind} and multiple delayed connections \cite{multd}, 
analytical study of AD under distributed delay \cite{ddnew1,ddnew2}, non-autonomous dynamics of Van der Pol oscillator  in AD regime \cite{nonauto}, 
coexistence of AD and synchronous oscillatory behavior in time delay systems \cite{co} and
controlling delay induced AD by coupling phase \cite{phase}.

Very recently Sekikawa and coworkers \cite{sudden} have explained the sudden transition from chaotic motion to the state of AD, a kind of 
transition observed earlier \cite{flip1,garima,amritkar2} too. 
The group has done the analysis for Bonhoeffer van der Pol oscillator and have found that such a direct 
transition to AD is a result of saddle-node bifurcation.

Study on AD has also been extended to networks wherein the works include exploring the effect of spatial distributions on AD \cite{spatfq}, 
effect of gradient coupling on AD in network \cite{grad},  insensitive dependence of AD on network structures \cite{netad} and 
AD in networks of delay coupled delay oscillators \cite{hofener}. An important scenario which observes AD is the reactive coupling \cite{react}. 
It increases the critical coupling strength for onset of AD  but when present alongside the meanfield interaction it supports the occurrence 
of AD by making the number of dead oscillators increase gradually in the network.

\section{The Phase-flip transition}

An interesting phenomenon that is frequently observed in coupled nonlinear systems within the regime of synchronization is the phase--flip transition~\cite{flip1,flip2,flip3,flip4,flip5,flip6,flip7,flip8,flip9,flip10,hamiltonian}. The relative phases of oscillations of the sub-systems change abruptly, typically by  $\pi$, when a parameter such as the  time--delay is varied. This phase change is accompanied by a change in frequency, and the transition is also quite general in the sense that it is observed in limit cycle as well as in chaotic oscillators, and furthermore the dynamics in the synchronized state can be periodic, quasiperiodic, or chaotic~\cite{flip2,flip3}.

While this is not a bifurcation since the largest Lyapunov exponent does not vanish, there is nevertheless an interesting feature of this transition, namely an (avoided) {\it crossing} of Lyapunov exponents \cite{flip4,flip8}. When this happens within the AD region all the exponents are strictly negative, and a detailed analysis can be carried out. Note however, that the dynamics, such as it is, is transient since the systems eventually settle onto fixed points. Thus the phase--flip here refers to the fact that there is a transition in the decaying dynamics of both the subsystems from being in phase to being out of phase. 

\subsection{Symmetric delay}

Consider the specific example of identical Landau--Stuart oscillators coupled with delayed interactions. The equation of motion is given by, in usual notation,
\begin{equation}
\dot{Z_i} = \left(1+ i\omega - |Z_i|^2 \right) Z_i+\epsilon[\left( Z_j(t-{\tau})- Z_i \right)];~~i, j = 1,2;~~i \ne j, \nonumber
\label{eq:ls-osc}
\end{equation}\noindent
and we have taken the amplitude to be unity. 
\begin{figure}
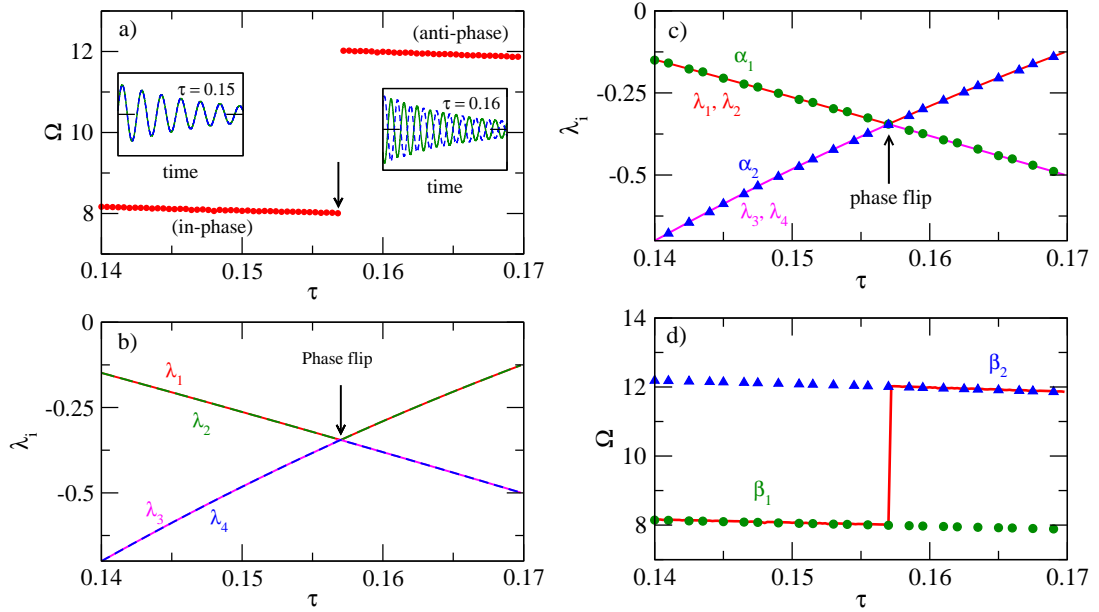

\begin{minipage}[b]{0.45\linewidth}
\centering
\scalebox{0.52}{\includegraphics{figure4ab.eps}}
\end{minipage}
\hspace{0.25cm}
\begin{minipage}[b]{0.45\linewidth}
\centering
\scalebox{0.54}{\includegraphics{figure4cd.eps}}
\end{minipage}
\caption{(Colour online) (a) And (b) show respectively the frequency jump and the crossings of the Lyapunov exponents for Landau-Stuart oscillators in AD regime. The insets in subfigure (a) are the transient dynamics of $x_1$ and $x_2$ (x variables of two oscillators) at two different delay values-- before ($\tau=0.15$) and after ($\tau=0.16$) the phase flip. In (b), the Lyapunov exponents are plotted by solid red ($\lambda_1$), dashed green ($\lambda_2$), solid magenta ($\lambda_3$) and dashed blue ($\lambda_4$) lines. (c) and (d) show the variation of real and imaginary parts of eigenvalues of the system 
(Eq.(\ref{eq:ls-osc})) with time delay. In (c), the real parts of the complex conjugate eigenvalue pair $\alpha_1$, $\alpha_2$ (green circles and blue triangles respectively) match with the Lyapunov exponents (solid lines). The corresponding imaginary parts $\beta_1$, $\beta_2$ are plotted in (d) with numerically calculated frequencies (solid line). The system parameters are $A=1, \omega=10.0, \epsilon=2.0$.}
\label{fig:pf1}
\end{figure}
Transforming into polar coordinates: $R_{1,2} = \sqrt{x_{1,2}^2+ y_{1,2}^2}$; $\theta = \tan^{-1}\left({y_{1,2}}/{x_{1,2}}\right)$ and assuming that the amplitudes vary slowly, the phase dynamics of the system is given by:
\begin{equation}
\dot{\theta_i} = \omega + \epsilon \sin \left[ \theta_j(t-{\tau}) - \theta_i(t) \right];~~i,j = 1,2;~~i \ne j.
\label{eq:km}
\end{equation}\noindent
In the synchronized region there is  a common frequency of oscillation which we denote $\Omega$ (in AD, this is the frequency of damped oscillations). If the phase difference between the subsystems is $\Delta \phi$, it can easily be shown~\cite{flip4} that this can either be zero or $\pi$, and the corresponding frequencies of oscillation satisfy the transcendental equations
\begin{eqnarray}
\Omega &=& \omega - \epsilon \sin(\Omega \tau), ~~ \mbox{for in-phase solutions}\\
\Omega &=& \omega + \epsilon \sin(\Omega \tau), ~~ \mbox{for anti-phase solutions}
\label{eq:freq}
\end{eqnarray}\noindent
At the parameter value where the phase--flip occurs there is a sudden change in synchronized frequency, Fig.~\ref{fig:pf1}(a), while in the spectrum of Lyapunov exponents  there is an avoided crossing between the two largest exponents and the next two, Fig.~\ref{fig:pf1}(b). (Since Lyapunov exponents are ordered by rank, they cannot in principle cross one other.) 

In the neighborhood of the transition, the Jacobian matrix at the fixed point (here the origin) has complex eigenvalues, and since this is a regime of AD, the real parts yield the Lyapunov spectrum.  Linearizing Eq.~(\ref{eq:ls-osc}) around the origin gives the characteristic equation
\begin{equation}
\mathrm{Det}(\mathbf{J} - \lambda \mathbf{I}) = 0,
\end{equation}\noindent
where $\mathbf{I}$ is the identity matrix and the $\mathbf{J}$ is the Jacobian. Assuming the perturbation varies as $e^{\lambda t}$, this yields
\begin{equation}
\lambda^2 - 2(a+i\omega)\lambda+(a^2-\omega^2 + i 2 a \omega)-\epsilon^2 e^{-2\lambda \tau}=0
\label{eq:char}
\end{equation}\noindent
where $a = 1 - \epsilon$. Substituting $ \lambda = \alpha + i \beta$ gives the pair of equations
\begin{equation}
\left.\begin{array}{l}
\alpha^2 - \beta^2 -2(a \alpha - \beta \omega) + a^2 - \omega^2 - \epsilon^2 e^{-2\alpha\tau} \cos 2\beta\tau = 0 \\
2 \alpha \beta -2(\alpha \omega + a \beta) + 2 a \omega + \epsilon ^2 e^{-2 \alpha \tau} \sin 2\beta\tau = 0\\
\end{array}\right\}
\label{eq:lcy10}
\end{equation}\noindent
which can be solved numerically to obtain $\alpha$ and $\beta$. 

As can be seen in Fig.~\ref{fig:pf1}(c)-(d), the real parts of the eigenvalues give the Lyapunov exponents,  and the imaginary part of the largest eigenvalue is the frequency of the damped oscillation for the coupled system. The eigenvalues come in complex conjugate pairs,  $\alpha_1 \pm \iota \beta_1$ and $\alpha_2 \pm \iota \beta_2$,  resulting in the spectrum of exponents having the degeneracies seen in Fig.~\ref{fig:pf1}(c). When the exponents cross, the imaginary part of the largest eigenvalue pair exchange their {\it imaginary} parts ($\beta_1, \beta_2$). This results in the phase and the frequency jump that are observed at the phase flip transition.  Similar behavior near the transition is found also for coupled chaotic oscillators in AD region~\cite{flip4}. 

\subsection{Asymmetric delay}

\begin{figure}[b]
\begin{minipage}[b]{0.45\linewidth}
\centering
\scalebox{0.23}{\includegraphics[angle=0]{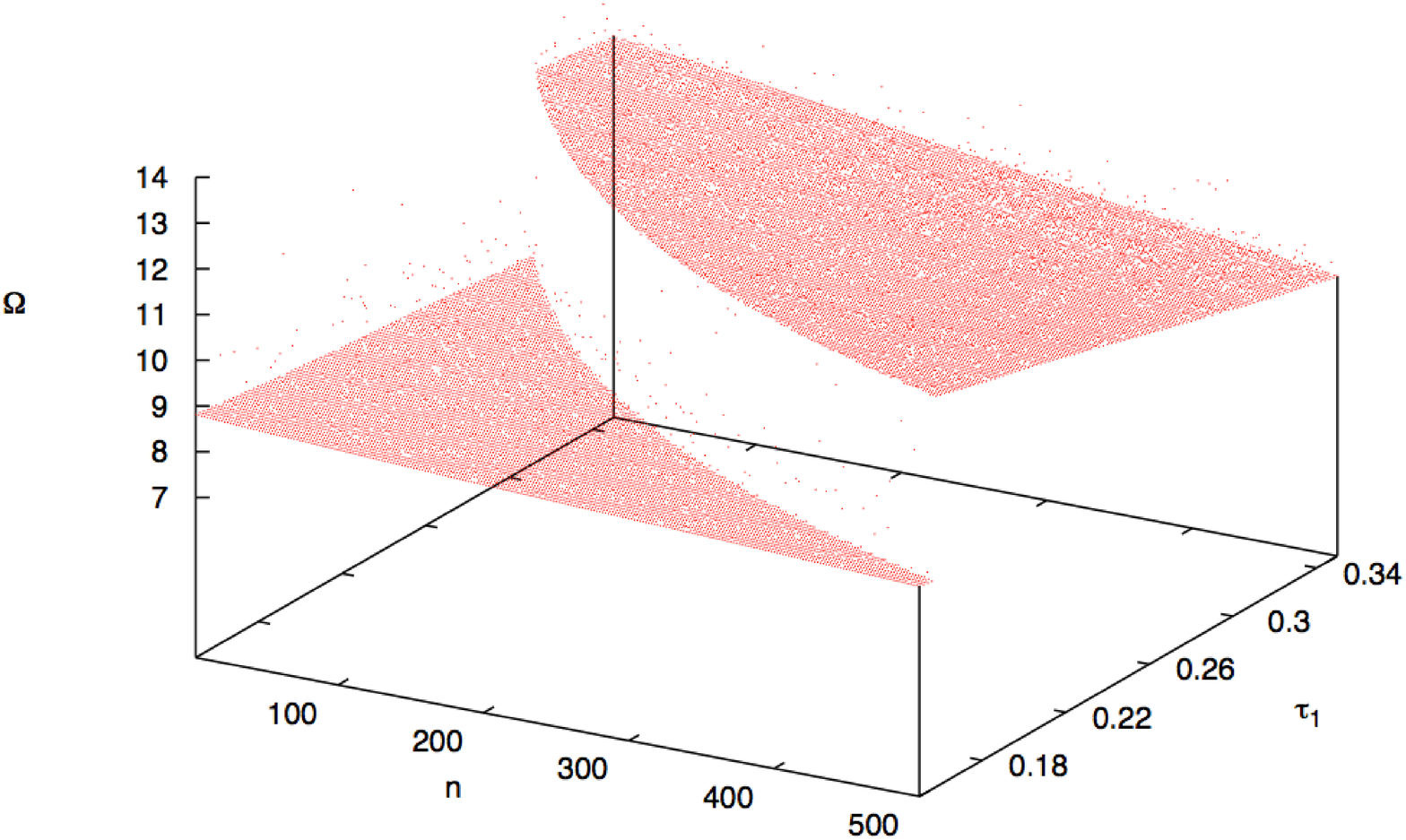}}
\end{minipage}
\hspace{0.5cm}
\begin{minipage}[b]{0.45\linewidth}
\centering
\scalebox{0.23}{\includegraphics[angle=0]{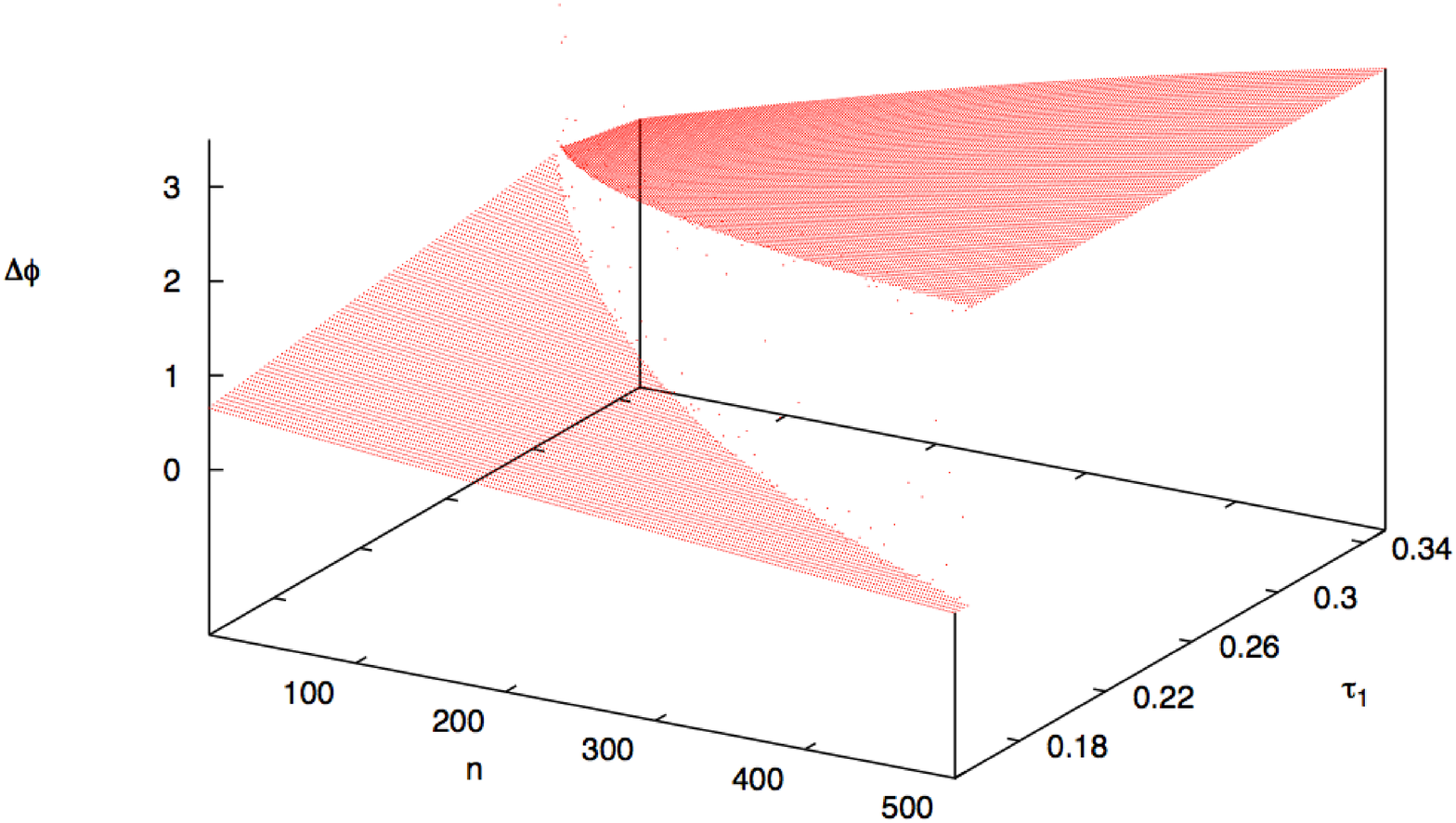}}
\end{minipage}
\caption{(Colour online) The variation of the numerically calculated synchronized frequency $\Omega$ as a function of delays $\tau_1$ and $\tau_2$. 
$\tau_2 = (n/N)\tau_1$ where $N = 500$ is the discretization taken in the simulation (left). 
The variation of the phase difference ($\Delta \phi$) as a function of delays $\tau_1$ and $\tau_2$ (right). 
System parameters are fixed as $\omega = 10, \epsilon = 2.0$.}
\label{fig:pf2}
\end{figure}

Phase-flip is also observed in oscillators coupled by asymmetric delays~\cite{flip8}, the case when the speed of information transmission is direction dependent. Consider the system of Landau--Stuart oscillators
\begin{equation}
\begin{array}{l}
\dot{Z_i} = (1 + \iota \omega-|Z_i|^2)Z_i +\epsilon [Z_j(t-{\tau_{i}})- Z_i];~~i,j = 1,2;~~i \ne j,
\end{array}
\label{eq:dd-ls}
\end{equation}\noindent with $\tau_1 \neq \tau_2$.
Proceeding as before, the synchronized frequencies in this case can be shown to be given by the zeros of the functions%
\begin{equation}
F_{\mp}({\Omega}) = {\omega} - {\Omega} \mp K \sin({\Omega}{\bar{\tau}})
\label{eq:ddfreq}
\end{equation}
where $\bar{\tau} = (\tau_1 + \tau_2)/2$ is the average delay, and the phase difference is given by~\cite{flip8}
\begin{equation}
\left.
\begin{array}{lll}
{\Delta \phi}  &= -\dfrac{{\Omega}{\Delta \tau}} {2} & \mbox{if~~} \cos \Omega \bar{\tau} > 0\\
\\
          &= \pi - \dfrac{{\Omega}{\Delta \tau}} {2} & \mbox{otherwise.} \\
\end{array}
\right\rbrace
\label{eq:ddphase}
\end{equation}
Although the phase difference differs from zero or $\pi$ and depends upon the difference of the individual delays, $\Delta \tau = (\tau_1 - \tau_2)$, the phase jump in this case is also accompanied by a discontinuity in synchronized frequencies as shown in Fig.~\ref{fig:pf2} where $\Omega$ and $\Delta$ computed numerically as a function of $\tau_1$ and $\tau_2$ are depicted. A line of phase difference discontinuity as well as the frequency jump can be seen.

The dependence of $\Omega$ on the arithmetic mean of the asymmetric delays has the interesting consequence that systems with the same average delay 
have identical the frequency responses. The eigenvalues of the Jacobian matrix at the fixed point also have the same property,%
\begin{equation}
\textbf{J} =
\left(
\begin{array}{cc}
{1+\iota \omega -\epsilon} & {\epsilon e^{-\lambda \tau_1}}\\
{\epsilon e^{-\lambda \tau_2}} & {1+\iota \omega -\epsilon}\\
\end{array}
\right)
\label{jaco-ls}
\end{equation}
since from Eq.(\ref{jaco-ls})  
one obtains the characteristic equations:

\begin{eqnarray}
\lambda^2 - 2(a+i\omega)\lambda+(a^2-\omega^2 + i 2 a \omega)-\epsilon^2 e^{-\lambda(\tau_1 + \tau_2)}=&0\\
\Rightarrow \lambda^2 - 2(a+i\omega)\lambda+(a^2-\omega^2 + i 2 a \omega)-\epsilon^2 e^{-2\lambda(\bar{\tau}) }=&0
\label{eq:dd-char}
\end{eqnarray}\noindent
where $a = (1 - \epsilon)$. Since Eq.~(\ref{eq:dd-char}) is a function of the average delay and does not separately depend upon the individual delays, the frequencies, eigenvalues and consequently the Lyapunov exponents in the AD region must be equal for systems with the same average delays, and  the asymmetric delay case is exactly equivalent to a symmetric system with the same average delay~\cite{flip8}. Here also there is an avoided crossing at the flip transition, with the exchange of imaginary parts of eigenvalue pairs~\cite{flip8}.

This characteristic of the phase flip transition, namely the avoided crossing of Lyapunov exponents along with the exchange of imaginary parts of the complex--conjugate eigenvalue pairs at the point, causing frequency and phase jumps have been observed in various periodic and chaotic systems~\cite{flip4,flip8}. Eigenvalue analysis within the AD regime is possible since all exponents are negative. When the dynamics is oscillatory, (numerical) experimental results indicate that there are crossings in the Lyapunov exponents, but the analysis for periodic or chaotic states is nontrivial. It seems likely, however that the phase--flip would have a similar mechanism in these cases~\cite{flip2,flip3,flip4,flip6,flip8,flip9} also. It would be interesting to extend the analysis that has been possible for AD to other dynamical states since 
the phase--flip transition is frequently seen in the dynamics of the systems coupled with or without delay---electronic circuits \cite{flip5}, neurons \cite{flip6}, electrochemical cells \cite{flip9}, plastic bottle oscillators \cite{flip5}.

\section{ Coupled Hamiltonian systems}
The possibility of analogous phenomena occurring in weakly dissipative systems has been explored recently by examining 
Hamiltonian dynamical systems with velocity coupling \cite{hamiltonian}. 
This form of interaction makes the systems dissipative creating the possibility for the occurrence of AD.

\begin{figure}
\centering
\includegraphics[width=12cm]{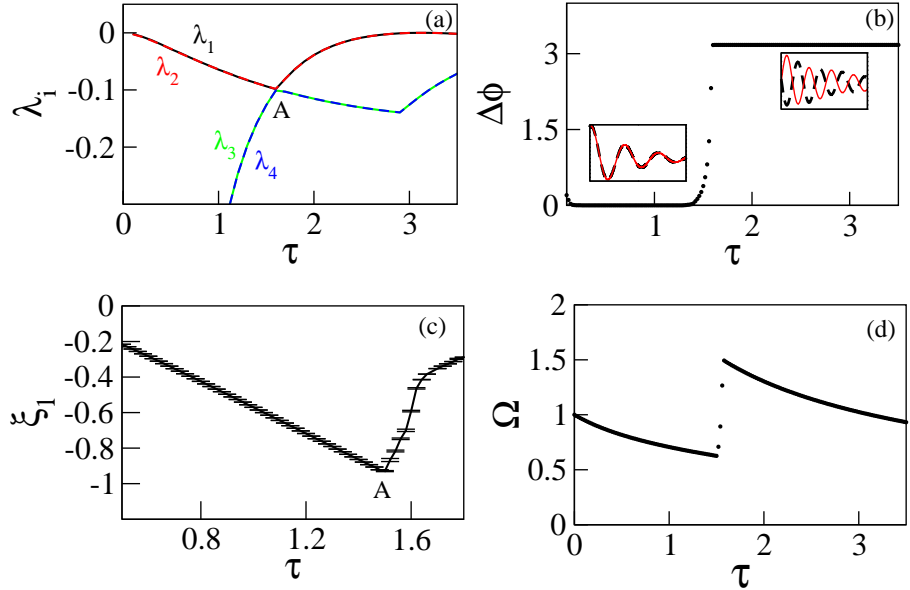}
\vspace{3cm}
\caption{The variation of (a) first four lyapunov exponents $(\lambda_i)$, (b) phase difference $(\Delta \phi)$ between the coupled oscillator, 
(c) energy dissipation rate $(\xi_{1})$ 
and (d) synchronized oscillation frequency $(\Omega)$ as function of time delay $\tau$.}
\label{fig:ho1}
\end{figure}

Consider a pair of simple harmonic oscillators,
\begin{eqnarray}
\ddot{x_{1}}+\omega^2_{1}=\epsilon(\dot{x_{2}}(t-\tau)-\dot{x_{1}}) \\
\ddot{x_{2}}+\omega^2_{2}=\epsilon(\dot{x_{1}}(t-\tau)-\dot{x_{2}}) \nonumber
\end{eqnarray}
$x_{1,2}$ and $\dot{x_{1,2}}$ are the positions and velocities of the two oscillators and  
$\omega_{1,2}$ their oscillation frequencies. 
Stability analysis around the fixed point $(x_{1,2},\dot{x}_{1,2})=(0,0)$ shows that occurrence of AD is independent of the  coupling strength, $\epsilon$. It follows from here that as soon as 
the delay is switched on AD occurs irrespective of the value of coupling strength, as can be seen in the largest  Lyapunov exponent, $\lambda_1$ (see Fig.(\ref{fig:ho1}a)). 
Note also that $\lambda_1 \to$ 0 at certain critical delays which are points of marginal stability, 
when the real part of the eigenvalues of Jacobian is zero. These points can be estimated  to be $\tau_{c}=nT/2$; here the coupling effectively vanishes,  and the systems decouple showing conservative dynamics again.  

The region of AD is again seen in the spectrum of Lyapunov exponents, Fig.~\ref{fig:ho1}a and  
here too transients show the phase--flip transition. The phase difference between the oscillators is plotted in Fig.(\ref{fig:ho1}b). In addition, since the uncoupled systems are Hamiltonian in nature, one can analyze the rate of energy decay \cite{eng}.  The energy of each oscillator is
\begin{eqnarray}
E_{1,2}=\frac{1}{2}(\dot{x_{1,2}}^2+\Omega^{2}x_{1,2}^2)
\label{eq:eng}
\end{eqnarray}
where $\Omega$ is the common frequency of oscillation. From the decaying maxima of energy 
$E_{1,2}^{m}$ ($m$ labels the successive maxima) one computes 
\begin{eqnarray} 
e_{1,2}&=\langle\,\log|E^{m=1}_{1,2}-E^{m}|\rangle_{m} \nonumber
\end{eqnarray}
from which one can obtain
\begin{eqnarray}
\xi_{1,2}&=\langle\,e_{1,2}\,\rangle
\label{eq:decayc}
\end{eqnarray}
where $\langle~~\rangle$ indicates an averaging over initial conditions. $\xi$ quantifies the rate of energy dissipation and a
plot of $\xi_{1}$ as function of the delay in the AD region is shown in Fig.~\ref{fig:ho1}c: energy dissipates faster prior to the flip transition, and slower thereafter. 

\section{Summary}
We have presented detailed numerical as well as analytical discussion of  amplitude death in coupled nonlinear systems. We have considered some special configurations \eg time--varying delay, environmental coupling and mean field coupling. 
Time-varying delays have advantages over fixed and distributed delay and are more realistic
in practical applications.  We have also considered environmental coupling, and we believe that
this form of interaction is ubiquitous in a wide range of systems, particularly in biology. Such 
coupling leads to a large regime of AD.  With feedback from a mean field, AD becomes possible in systems coupled by identical variables even in the absence of time delay. 
We also considered coupled Hamiltonian systems with time--delay coupling leading to AD.

Both in--phase and out--of--phase dynamics is known to be stabilized in synchronous systems \cite{earl} and a transition between these is known in a variety of situations \cite{nirmal}. The phase--flip transition can be analyzed in details in the regime of AD, both for the case of symmetric delays as well as when the delay is direction dependent.  

Amplitude death can be a  crucial and desired feature in a variety of fields, and thus mechanisms for achieving AD  can find application in different areas. This paper extends our recent review \cite{review}  wherein the important coupling schemes and scenarios that lead to oscillation death were 
discussed.  The ability to obtain a specific fixed point, namely the process of targeted amplitude death is important from an engineering point of view. Finally, this analysis can also be instructive in achieving the opposite objective, namely the avoidance of AD. Taken together, these studies are therefore important in the context of controlling the dynamics of nonlinear systems.

\section{acknowledgements}

We have great pleasure in dedicating this article to Abhijit Sen in appreciation of many years of his academic leadership, friendship and generosity. His seminal contributions to the study of amplitude death phenomena sparked our interest in the area, and our continued interaction with him has been a great education.  GS has been supported by the CSIR through a Senior Research Fellowship, NP by the DST, AP acknowledges the financial support of Delhi University--DST PURSE programme, and RR the JC Bose fellowship.


\begin{thebibliography}{9}

\bibitem{hys} A. Prasad, L. D. Iassemidis, and S. Sabesan, {\it Pramana}, {\bf 64}, 513 (2005). 
\bibitem{rid} J. C. Sommerer and E. Ott, {\it Nature}, {\bf 365}, 136 (1993).
\bibitem{rajatphyscon} R. Karnatak, R. Ramaswamy and A. Prasad, {\it PHYSCON}, (2009)
\bibitem{kurth} A. Pikovsky, M. Rosenblum, J. Kurths, {\it Synchronization: A universal concept in 
Nonlinear Sciences}, (Cambridge University Press, 2001) 
\bibitem{review} G. Saxena, A. Prasad and R. Ramaswamy, {\it Phys. Rep.}, {\bf 521}, 205, (2012).
\bibitem{laser} K. Pyragus, F. Lange, T. Letz, J. Parisi and A. Kittel, {\it Phys. Rev. E}, {\bf 61}, 3721 (2000).
\bibitem{thermo} K. P. Zeyer, M. Mangold and E. D. G. Peng, {\it J. Phys. Chem.}, {\bf 105}, 7216 (2001).
\bibitem{mech} M. A. Barron, I. Hilerio and  G. Plascencia, {\it Adv. in Mech. Eng.}, {\bf 2012}, 746537 (2012).
\bibitem{power} S. R. Huddy and J. D. Sufca, {\it IEEE Trans.: Power electronics}, {\bf 28}, 247 (2013).
\bibitem{nd1} D. J. Selkoe, {\it Ann. NY Acad. Sci.} {\bf 924}, 17 (2000).
\bibitem{nd2} R. E. Tanzi, {\it Nat. Neurosci.} {\bf 8}, 977 (2005).
\bibitem{nd3} B. Caughey and P. T. lansbury, {\it Annu. Rev. Neurosci.}, {\bf 26}, 267 (2003). 
\bibitem{hear} K. H. Ahn, {\it arXiv:1202.5912}.
\bibitem{ocean} B. Gallego and P. Cesso, {\it J. Clim.}, {\bf 14}, 2815 (2001).

\bibitem{aronson} D. G. Aronson, G. B. Ermentrout, and N. Kopell, {\it Physica D}, {\bf 41}, 403 (1990).
\bibitem{delay1} D. V. R. Reddy, A. Sen and G. L. Johnston, {\it Phys. Rev. Lett.}, {\bf 80}, 5109 (1998).
\bibitem{delay2} S. H. Strogatz, {\it Nature}, {\bf 394}, 316 (1998).

\bibitem{cc}R. Karnatak, R. Ramaswamy, and A. Prasad, \pre{76}{035201}{2007}.
\bibitem{dc}K. Konishi,  \pre{68}{2003}{067202}.
\bibitem{dcall}K. Konishi, {\it Int. J. Bifurcation Chaos}, {\bf 17}, 2781 (2007); 
K. Konishi and N. Hara, {\it Phys. Rev. E}, {\bf 83}, 036204 (2011).
\bibitem{nlc}A. Prasad, M. Dhamala, B. M. Adhikari, and R. Ramaswamy, \pre{81}{027201}{2010}.
\bibitem{la}P. R. Sharma, A. Sharma, M. D. Shrimali, and A. Prasad, \pre{83}{067201}{2011}.
\bibitem{atay} F. M. Atay, {\it Phys. Rev. Lett.}, {\bf 91}, 094101 (2003).
\bibitem{garima} G. Saxena, A. Prasad and R. Ramaswamy, {\it Phys. Rev. E}, {\bf 82}, 017201 (2010).
\bibitem{konishi} K. Konishi, H. Kokame and H. Hara, {\it Phys. Lett. A}, {\bf 374}, 733 (2010).
\bibitem{amritkar} V. Resmi, G. Ambika and R. E. Amritkar, {\it Phys. Rev. E}, {\bf 81}, 046216 (2010).
\bibitem{meanf} A. Sharma and M. D. Shrimali, {\it Phys. Rev. E}, {\bf 85}, 057204 (2012).
\bibitem{grad} W. Liu, J. Xiao, L. Li, Y. Wu and M. Lu, {\it Nonlinear Dyn.}, {\bf 69}, 1041 (2012).
\bibitem{partial}W. Zou and M. Zhan, {\it Phys. Rev. E}, {\bf 80}, 065204 (2009).
\bibitem{indirect}A. Sharma, P. R. Sharma and M. Shrimal, {\it Phys. Lett. A}, {\bf 376}, 1562, (2012).

\bibitem{konishi2} Y. Sugitani, K. Konishi and N. Hara, {\it Nonlinear Dyn.}, {\bf 70}, 2227 (2012).

\bibitem{amritkar2} V. Resmi, G. Ambika and R. E. Amritkar, {\it Phys. Rev. E}, {\bf 84}, 046212 (2011). 
\bibitem{amritkar3} V. Resmi, G. Ambika, R. E. Amritkar and G. Rangarajan, {\it Phys. Rev. E}, {\bf 85}, 046211 (2012).

\bibitem{diffc} J. K. Hale, {\it J. Dyn. and Differ. Equ.}, {\bf 9}, 1, (1997).
\bibitem{ap} A. Prasad, {\it to be published}.

\bibitem{asym1} W. Zou, X. G. Wang, Q. Zhao and M. Zhan, {\it Front. Phys. China}, {\bf 4}, 97 (2009).
\bibitem{asymeps} W. Zou, Y. Tang, L. Li and J. Kurths, {\it Phys. Rev. E}, {\bf 85}, 046206 (2012).
\bibitem{rep} Y. Chen, J. Xiao, W. Liu, L. Li and Y. Yang, {\it Phys. Rev. E}, {\bf 80}, 046206 (2009).
\bibitem{orderpara} C. Yao, W. Zou and O. Zhao, {\it Chaos}, {\bf 22}, 023149 (2012).
\bibitem{sind} L. B. Le, K. Konishi and N. Hara, {\it PHYSCON} (2012).
\bibitem{multd} L. B. Le, K. Konishi and N. Hara, {\it NESD- Nonlinear Dynamics of electronic systems: 
Conference Proceedings} (2012).
\bibitem{ddnew1} Y. N. Kyrychko and K. B. Blyuss, {\it Eur. Phys. J. B.}, {\bf 84}, 307 (2011)
\bibitem{ddnew2}Y. N. Kyrychko, K. B. Blyuss and E. Scholl, {\it arXiv}, 1209.0133.

\bibitem{nonauto} A. P. Kuznetsov, E. P. seleznev and N. V. Stankevich, {\it  Comm. Nonlinear Sc. and Num. Sim.}, {\bf 17}, 3740 (2012).
\bibitem{co} L. S. Jin, L. Ying, S. Lu and Y. Zhang, {\it Multimedia \& Signal Processing International Conference}, {\bf 2}, 21 (2011).
\bibitem{phase} W. Zou, J. Lu, Y. Tang, C. Zhang and J. Kurths, {\it Phys. Rev. E}, {\bf 84}, 066208 (2011).
\bibitem{sudden} M. Sekikawa, K. Shimizu, N. Inaba, H. Kita, T. Endo, K. Fujimoto, T. Yoshinaga and K. Aihara, {\it Phys. Rev. E}, {\bf 84}, 056209 (2011).
\bibitem{flip1} A. Prasad, \pre{72}{056204}{2005}.
\bibitem{spatfq} Y. Wu, W. Liu, J. Xiao, W. Zou and J. Kurths, {\it Phys. Rev. E}, {\bf 85}, 056211 (2012).
\bibitem{netad} W. Zou, X. Zheng and M. Zhan, {\it CHAOS}, {\bf 21}, 023130 (2011).
\bibitem{hofener} J. M. Hofener, G. C. Sethia and T. Gross, {\it arXiv}, 1210.2002. 
\bibitem{react} W. J. Hua and L. X. Wen, {\it Chinese Phys. Lett.}, {\bf 26}, 030505 (2009).

\bibitem{flip2} A. Prasad, J. Kurths, S. K. Dana and R. Ramaswamy, \pre{74}{035204(R)}{2006}.
\bibitem{flip3} A. Prasad, S. K. Dana, R. Karnatak, J. Kurths, B. Blasius, and R. Ramaswamy, \CHAOS {\bf 18}, 023111 (2008).
\bibitem{flip4} R. Karnatak, N. Punetha, A. Prasad and R. Ramaswamy, \pre{82}{046219}{2010}.
\bibitem{flip5}  L. M. Cruz, J. Escalona, P. Parmananda, R. Karnatak, A. Prasad and R. Ramaswamy, \pre{81}{046213}{2010}.
\bibitem{flip6} B. M. Adhikari, A. Prasad, M. Dhamala, {\it CHAOS} {\bf 21}, 023116 (2011).
\bibitem{flip7}  A. Sharma, M. D. Shrimali, A. Prasad, R. Ramaswamy and U. Feudel, \pre{84}{016226}{2011}.
\bibitem{flip8}  N. Punetha, R. Karnatak, A. Prasad, J. Kurths and R. Ramaswamy, \pre{85}{046204}{2012}.
\bibitem{flip9}  A. Sharma, M. D. Shrimali and S. K. Dana, \CHAOS {\bf 22}, 023147 (2012).
\bibitem{flip10}  M. Kohira, H. Kitahara, N. Magome and K. Yoshikawa, \pre{85}{026204}{2012}.

\bibitem{hamiltonian} G. Saxena, A. Prasad and R. Ramaswamy, \pre{}{}{2013}--In press.
\bibitem{eng} Z. Wang and H. Hu, {\it Proceedings of International Design
Engineering Technical Conferences \& Computers and Information in Engineering Conference} {(2005)}.
\bibitem{earl} M. G. Earl and S. H. Strogatz, {\it Phys. Rev. E}, {\bf 67}, 036204 (2003). 
\bibitem{nirmal} N. Punetha and R. Ramaswamy {\it in progress}.

\end{thebibliography}
\end{document}